\documentclass[journal]{IEEEtran}
\usepackage{amsmath,amssymb}
\usepackage{booktabs}
\usepackage{url}
\usepackage{graphicx}
\usepackage{placeins}
\usepackage{float}
\usepackage[hidelinks]{hyperref}
\hypersetup{pdftitle={SpeakerTurn: Target-Speaker Semantic Turn Detection in Multi-Speaker Environments},pdfauthor={Nan Li, Hetao Shan, Jintao Liang, Jiahao Yan, Longbiao Wang, Tubin Huang, Min Xu},pdfsubject={Target-speaker semantic turn detection},pdfkeywords={speech, target speaker, turn taking, semantic turn detection}}
\title{SpeakerTurn: Target-Speaker Semantic Turn Detection in Multi-Speaker Environments}
\author{Nan Li$^{1}$, Hetao Shan$^{1}$, Jintao Liang$^{2}$, Jiahao Yan$^{2}$, Longbiao Wang$^{2}$, Tubin Huang$^{3}$, and Min Xu$^{3}$\\[0.5em]
\normalsize $^{1}$School of Computer and Information Engineering, Tianjin Normal University, Tianjin, China\\
\normalsize $^{2}$School of Artificial Intelligence, Tianjin University, Tianjin, China\\
\normalsize $^{3}$Shanghai Shenqijiyuan Intelligent Terminal Co., Ltd.}
\begin{document}
\maketitle
\raggedbottom
\begin{abstract}
Semantic completeness alone cannot determine when an assistant should respond in a multi-speaker environment: a completed utterance may belong to an interfering speaker, while the designated user is still speaking, acknowledging, or requesting time. We therefore study semantic turn inference conditioned on speaker identity. We propose SpeakerTurn, a framework that jointly predicts the target transcript and turn state from a speech mixture and an independent target-speaker reference. Built on CocktailASR-1, it links semantic decisions to the corresponding target content through joint supervision, distinguishing complete, incomplete, backchannel, wait, and target-absent states, with a separate interruption task. We also examine whether structured chain-of-thought (CoT) supervision improves decisions by representing speaker presence and semantic cues. On a synthetic multi-speaker test set, SpeakerTurn achieves five-state macro F1 of 0.811 and target-transcription character error rate (CER) of 4.20\%, correctly resolving all ten reference-swapped interruption pairs. CoT reduces transcription error but lowers turn macro F1 on both validation and test data. These results highlight the need to assess target attribution, semantic turn discrimination, and transcription jointly when adapting speech language models for speaker-conditioned interaction.
\end{abstract}
\section{Introduction}
A conversational system must determine whose speech matters and whether that speaker has finished. In a multi-speaker environment, an interfering speaker may complete a sentence while the designated user pauses midway through an explanation. Even a complete request to wait may retain the floor. SpeakerTurn addresses this ambiguity by conditioning semantic turn inference on a target-speaker reference.

Speaker-conditioned activity estimation and semantic turn prediction address complementary parts of this problem. Personal voice activity detection and target-speaker diarization associate acoustic activity with a specified speaker~\cite{ref190804284,ref220403793,ref200507272,ref240108052}. Acoustic and linguistic turn models estimate whether a contribution supports completion or continuation~\cite{ref240104868,ref260810878,ref260401897,ref260326515,ref250920410,smartturn}. Easy Turn distinguishes complete, incomplete, backchannel, and wait states~\cite{easyturn}, while SoulX-Duplug addresses streaming state prediction~\cite{soulx}. Interruption detection further distinguishes an intended intervention from other overlapping speech~\cite{ref260324144}. Semantic decisions must therefore remain associated with the enrolled speaker despite competing interpretations.

\begin{figure*}[!t]
\centering
\includegraphics[width=.94\textwidth]{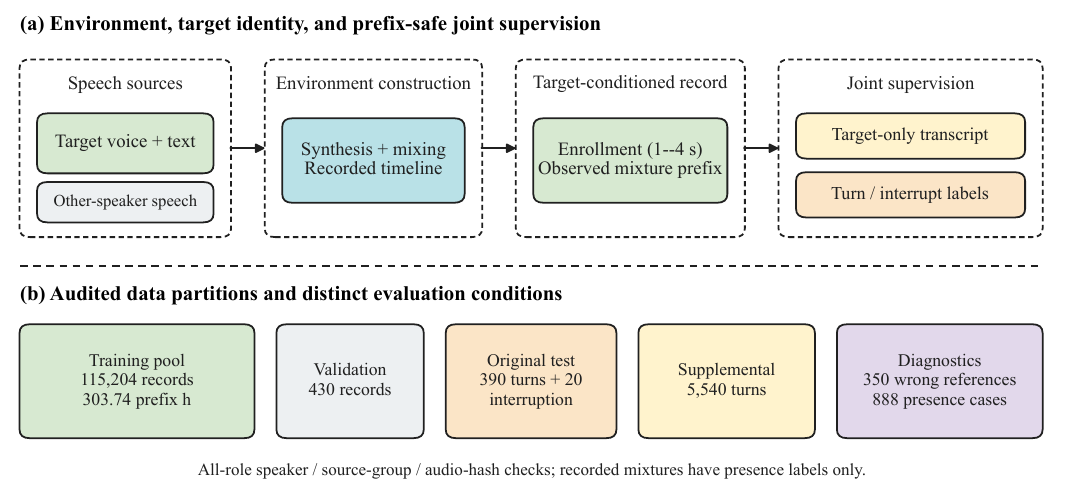}
\caption{Data construction and evaluation partitions. Supervision uses the observed prefix and synthetic generation receipts. Recorded-voice mixtures carry presence labels only. The experiments below focus on the original synthetic test and the matched validation comparison.}
\label{fig:data}
\end{figure*}

Full-duplex systems place this attribution problem within continuous listening and generation. Modular and speech-language architectures model their interaction~\cite{ref220515060,ref251218706,ref250908753,ref241000037,ref260206053}, and synchronous language models support simultaneous input and output~\cite{ref240802622,ref240915594,ref240615718,ref250517060,ref260706540,ref260614528}; a survey reviews this design space~\cite{ref250914515}. Benchmarks assess overlap, interruption, timing, and multi-turn behavior~\cite{ref260421406,ref250304721,ref250723159,ref251007838,ref260404847,ref260825218,ref250301174,ref250926388}. We focus on the speaker-specific semantic decision: which words belong to the designated user, and what conversational state do they express? Transcription, turn discrimination, and interruption attribution provide distinct evidence.

Pretrained speech representations and recognition models connect acoustic observations to linguistic content~\cite{ref200611477,ref221204356}. CocktailASR-1 provides a suitable foundation because it transcribes a target directly from a mixture using reference speech and supports empty output when the target is absent~\cite{cocktail}. We extend this interface by jointly supervising the target transcript and semantic state, making the relevant speaker's words an explicit learning target. Synthetic dialogue and audio-construction methods enable controlled comparisons~\cite{ref260726178,ref260325750}, subject to the limits of generated annotations and known observation boundaries.

We propose SpeakerTurn, a reference-conditioned large language model (LLM) that jointly predicts target content and semantic turn state, with a separate target-speaker interruption task. A structured chain-of-thought (CoT) variant additionally represents target presence, semantic cues, and assistant context. This tests intermediate supervision without assuming faithful internal reasoning.

The main contributions are as follows:
\begin{itemize}
\item We formulate target-conditioned semantic turn inference with five states and a separate interruption decision.
\item We adapt CocktailASR-1 through joint target-transcript and turn supervision, comparing decision-only and structured-CoT configurations.
\item We evaluate turn discrimination, target transcription, and recognition retention separately, revealing that strong performance in one does not ensure the others.
\end{itemize}

\section{SpeakerTurn}
\subsection{CocktailASR-1 Backbone and Inherited Capabilities}
CocktailASR-1 performs reference-conditioned automatic speech recognition (ASR) directly on a speech mixture, without an explicit separation stage~\cite{cocktail}. Its released interface supports single-speaker recognition, empty transcripts for absent targets, and an optional native recognition-CoT mode. SpeakerTurn builds on these capabilities through a joint transcription-and-turn interface. Native recognition CoT and the structured turn-CoT supervision used here are distinct interfaces.

\begin{figure*}[!t]
\centering
\includegraphics[width=.94\textwidth]{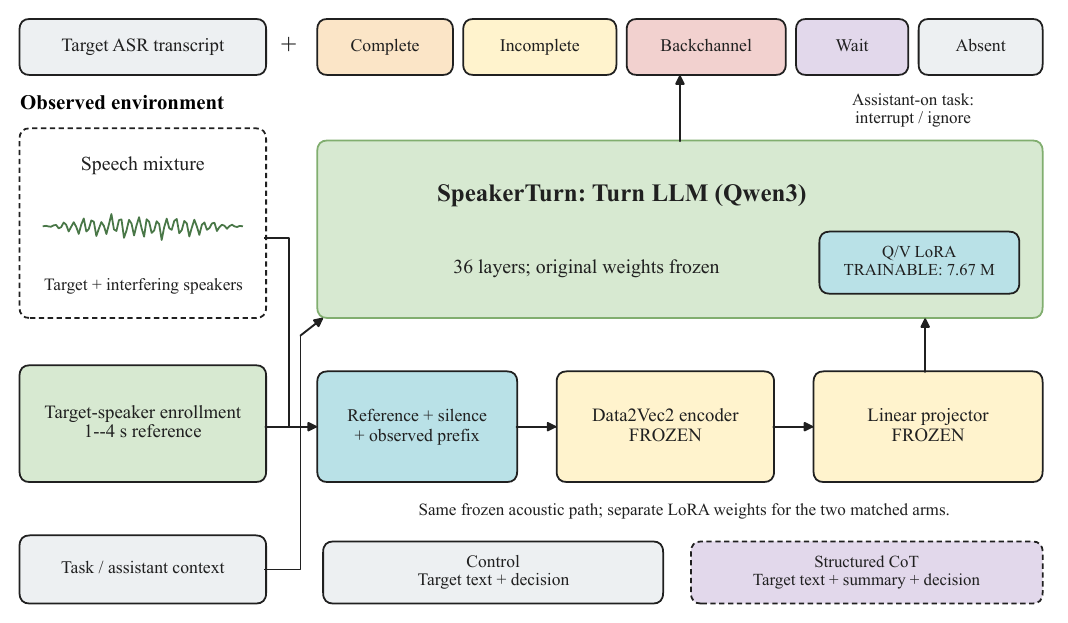}
\caption{Target-conditioned Turn LLM and adaptation scope. Both joint arms inherit the same frozen acoustic path and use separate LLM LoRA weights. Target text and decisions are generated outputs; a structured summary is added only in the CoT arm.}
\label{fig:architecture}
\end{figure*}

As shown in Fig.~\ref{fig:architecture}, the enrollment and observed mixture pass through the inherited Data2Vec2 acoustic encoder and linear projector to the Qwen3 language-model backend. Adaptation is restricted to low-rank adaptation (LoRA)~\cite{lora} parameters in the query/value projections: 144 tensors comprising 7,667,712 trainable parameters, with rank 16, scaling coefficient 32, and dropout 0.05. The acoustic encoder, projector, original decoder weights, and output head are frozen. Environmental context is thus provided through the mixture itself, without introducing a separate environment encoder, separation network, or turn-classification head.

\subsection{Problem Formulation}
Let $r$ denote the enrollment recording, $x_{1:t}$ the mixture observed up to time $t$, and $q$ the task instruction. SpeakerTurn models the target-conditioned output $y$ as
\begin{equation}
P_{\theta}(y\mid r,x_{1:t},q).
\end{equation}
The semantic turn task assigns one of five labels: complete, incomplete, backchannel, wait, or target absent. In the separate interruption task, the assistant is assumed to be speaking and the output is interrupt or ignore. Both tasks are conditioned on the enrollment, so the same observed intervention can require a different decision when the target reference changes.

\subsection{Joint Transcription and Turn Supervision}
The joint output consists of a target transcript enclosed in \texttt{<answer>} tags, followed by a state enclosed in \texttt{<decision>} tags. The structured-CoT variant prepends a \texttt{<think>} block containing target presence, a semantic cue, and assistant context. Both variants freely generate their outputs; they do not select from candidate label scores.

Joint training combines the mean token-level cross-entropy of each output segment:
\begin{equation}
\mathcal{L}=\mathcal{L}_{\mathrm{text}}+\mathcal{L}_{\mathrm{decision}}
+0.2\mathcal{L}_{\mathrm{CoT}}+0.1\mathcal{L}_{\mathrm{EOS}},
\end{equation}
The CoT term is omitted for the control model. Target transcripts are supervised outputs and are never supplied as inference inputs. Target-absent examples and non-target interventions use an empty answer. The auxiliary cues are deterministic templates derived from synthetic labels; they do not establish faithful internal reasoning. Native-ASR replay uses its separate transcription objective. Joint decoding stops at the closing decision tag, the official end-of-sequence (EOS) token, or the generation limit specified below.

\section{Experimental Setup}
\subsection{Data Construction and Tasks}
\textit{Text and semantic labels.} Qwen3-8B generates bilingual text plans, followed by two DeepSeek expansions using the \texttt{deepseek-flash} API in non-thinking mode. Prompts specify language, everyday topic, speaker roles, acoustic scenario, and semantic state. Incomplete turns omit required complements; backchannels acknowledge without new requests; wait utterances request pauses. Chinese/English versions share content groups and splits. Schema, language, role-order, duplicate-text, and DeepSeek-specific length/punctuation/pause checks constrain generation without independently validating labels.

\textit{Enrollment and acoustic construction.} AISHELL-1 and LibriSpeech provide Chinese and English speaker references, respectively. Each identity uses separate 1--4 s utterances for synthesis conditioning and enrollment. CosyVoice2-0.5B synthesizes planned utterances under up to five deterministic, language- and split-matched speaker assignments, with distinct target and interfering identities. Scenarios comprise target-only, non-target-only, two-speaker overlap, target-with-noise, and three-speaker speech. The target starts halfway through interfering speech in overlap; three-speaker sequences use 150 ms gaps with the target last. Gaussian-noise signal-to-noise ratios are 5, 10, 15, or 20 dB. Mixtures are peak-scaled at 16 kHz; room acoustics and device echo are not simulated.

\textit{Supervision and observation.} Inputs end at generator-defined target endpoints; transcripts supervise observed target speech only. Non-target-only inputs have absent labels and empty transcripts. Interruption commands use target and non-target enrollments to yield interrupt and ignore labels, respectively. An assistant-speaking flag provides context without playback in the waveform. Labels and text come from synthetic plans without human annotation or ASR verification. Known cutoffs define offline prefix decisions rather than continuous endpoint detection in human dialogue.

\textit{Training pool.} The 115,204 frozen records comprise 114,994 turn and 210 interruption conditions (57,709 Chinese, 57,495 English); DeepSeek expansions contribute 108,304. Observed prefixes span 303.74 h across 115,099 waveform hashes, excluding repeated presentations. This is retained training volume, not the planned synthesis volume.

\textit{Validation and original test.} These are distinct partitions of the same constructed corpus. Validation contains 430 conditions from 42 content groups and 74 speaker identities; the original test contains 410 conditions from 39 groups and 62 identities. For complete, incomplete, backchannel, wait, and absent, validation has 240, 40, 30, 40, and 70 conditions, respectively; test has 200, 60, 50, 40, and 40. Each turn class is equally divided between Chinese and English. Interruption adds 10 validation and 20 test conditions, all Chinese: five and ten command waveforms, respectively, each paired with target and non-target enrollments. Consequently, validation contains 220 Chinese and 210 English conditions, and test contains 215 Chinese and 195 English conditions. These counts represent evaluation conditions, with 425 and 400 distinct waveforms. Audits verify mutually disjoint content groups, waveform and enrollment hashes, and identities across all speaker roles among training, validation, and original test.

\textit{ASR and reduced-state subsets.} Target ASR and four-state detection each use 350 target-present turns per partition (175 per language): original test in Table~\ref{tab:asr_main}, disjoint validation in Table~\ref{tab:cot_main}. Test references contain 18,835 normalized characters and 3,535 English words; validation has 17,190 and 3,210. Binary complete/incomplete detection uses 260 test (200/60) and 280 validation (240/40) conditions. These tasks reuse subsets of each partition rather than adding records.

\textit{Supplemental evaluation.} An additional 5,540 synthetic turn conditions include 4,790 target-present and 750 absent examples. Content groups and waveforms are disjoint from training, validation, and original test; the 62 speaker identities are shared with the original test. It evaluates additional content with the original test speaker pool.

We will release the code and synthesis scripts.\footnote{Planned code and data-synthesis repository: \url{https://github.com/linan2/SpeakerTurn}.}

\subsection{Training and Decoding}
The two joint-training variants start from the same initial 1,000-update adapter. Each receives 164,278 joint presentations, consisting of one visit to every eligible record and 49,074 additional interruption presentations, together with 54,759 native-ASR replays. The resulting schedule contains 219,037 updates per variant. AdamW uses a learning rate of $10^{-5}$, batch size one, seed 20261028, and gradient clipping at 1. Initialization, record order, replay schedule, and final-checkpoint selection are identical between variants. Optimizer and random states are restored across training segments. Matching update counts controls the presentation schedule, but not the additional tokens and computation introduced by CoT.

The earlier turn-only model provides a baseline without target-transcript supervision. It uses the same 115,204-record pool and initial adapter, but is trained on decision labels for 164,278 updates at a learning rate of $10^{-4}$, without transcript loss or native-ASR replay. Its decisions use validation-calibrated candidate scoring. Both models receive enrollment and mixture audio. The comparison consequently measures two complete training configurations; changes in learning rate, replay, update count, and readout prevent attributing their differences solely to transcript supervision.

Joint inference uses greedy generation with a limit of 384 new tokens. Native-prompt ASR is decoded separately with a 256-token limit. Only the enrollment, observed mixture, and task instruction enter inference. Gold transcripts, decisions, and structured cues are reserved for supervision or scoring. The reported models use their fixed final checkpoints; validation results do not determine the full-pool checkpoint selection.

\subsection{Baselines and Metrics}
Easy Turn~\cite{easyturn}, evaluated with its official Qwen2.5 tokenizer, provides a four-state control without target enrollment. Its wait policy differs from ours. Frozen CocktailASR-1 uses validation-calibrated candidate-label scores, whereas SpeakerTurn generates a joint output. Comparisons retain these input and readout differences. Smart Turn~\cite{smartturn} provides a binary control without enrollment. Binary-only controls appear only in the complete/incomplete comparison.

Moshi~\cite{ref241000037} is evaluated without enrollment as a native response-timing proxy, with fresh state and greedy decoding for each prefix. After the known observation cutoff, we append 3\,s of silence. A qualifying native text token within 2\,s of the cutoff on the input clock indicates complete; otherwise, incomplete. The fixed token rule excludes control tokens and retains non-control whitespace tokens. This readout measures response timing rather than explicit turn classification or wall-clock endpoint latency.

FireRedASR2-AED~\cite{firered2}, Qwen3-ASR-1.7B~\cite{qwen3asr}, and Whisper large-v3~\cite{whisperv3} are evaluated on the same 350 observed mixtures without enrollment. Their original hypotheses are scored against target-only references, so non-target words remain errors even when recognized correctly. CocktailASR-1 and SpeakerTurn additionally receive the target reference. All systems use mixture audio; oracle target stems are excluded.

Turn performance is measured by accuracy (Acc.), the fraction of correct predictions, and macro F1, the unweighted mean of classwise F1 scores, for five states ($n=390$), four target-present states ($n=350$), and complete/incomplete ($n=260$; 200/60 conditions). Malformed outputs and predictions outside the evaluated subset remain errors. Character error rate (CER) is micro-averaged after Unicode NFKC normalization, case folding, and alphanumeric filtering. English word error rate (WER) uses Latin alphanumeric word tokens, preserving internal apostrophes. Both error rates are uncapped. Independent checks reconstruct the reported full-pool metrics from raw outputs and verify manifest coverage and final-checkpoint provenance.

\section{Results and Discussion}
\subsection{Target-Speaker Turn Detection}
Table~\ref{tab:turn_main} compares five-state, four-state, and binary turn detection. SpeakerTurn leads each panel, reaching five-state macro F1 of 0.811 and accuracy of 83.33\%. Joint training retains turn performance close to turn-only while supporting transcription. Differences in training schedules and readouts preclude isolating the effect of transcript supervision.

\begin{table}[!t]
\centering\small
\caption{Turn detection on identical test subsets.}
\label{tab:turn_main}
\setlength{\tabcolsep}{3pt}
\renewcommand{\arraystretch}{1.12}
\begin{tabular*}{\columnwidth}{@{\extracolsep{\fill}}lrr@{}}\toprule
Model & Macro F1 & Acc. (\%)\\\midrule
\multicolumn{3}{l}{\textit{(a) Five states: 390 conditions}}\\
CocktailASR-1, frozen & 0.197 & 49.49\\
Turn-only (no text) & 0.805 & 82.31\\
SpeakerTurn (no CoT) & \textbf{0.811} & \textbf{83.33}\\
SpeakerTurn (CoT) & 0.764 & 80.26\\\midrule
\multicolumn{3}{l}{\textit{(b) Four target-present states: 350 conditions}}\\
Easy Turn & 0.312 & 58.57\\
CocktailASR-1, frozen & 0.259 & 55.14\\
Turn-only (no text) & 0.769 & 80.29\\
SpeakerTurn (no CoT) & \textbf{0.779} & \textbf{81.43}\\
SpeakerTurn (CoT) & 0.717 & 78.00\\\midrule
\multicolumn{3}{l}{\textit{(c) Complete/incomplete: 260 conditions}}\\
Moshi$^{\dagger}$ & 0.396 & 39.62\\
Smart Turn & 0.579 & 68.46\\
Easy Turn & 0.658 & 78.08\\
CocktailASR-1, frozen & 0.507 & 72.69\\
Turn-only (no text) & 0.776 & 84.62\\
SpeakerTurn (no CoT) & \textbf{0.804} & \textbf{86.54}\\
SpeakerTurn (CoT) & 0.796 & 86.15\\
\bottomrule\end{tabular*}
\par\vspace{3pt}{\footnotesize\raggedright All methods within each panel use identical sample IDs, waveform prefixes, cutoffs, and gold labels. Four states exclude target absence. Easy Turn, Smart Turn, and Moshi receive no enrollment. Easy Turn uses a different wait policy. $^{\dagger}$Moshi uses the response-timing proxy defined in the evaluation protocol; it does not emit turn labels.\par}
\end{table}

On the binary subset, SpeakerTurn outperforms Easy Turn and Smart Turn. Moshi's score describes a response-timing proxy, with qualifying text also generated before every cutoff, and should be interpreted separately from explicit turn classification.

Errors concentrate on premature completion: 25/60 incomplete turns and 22/50 backchannels are predicted as complete. All 40 target-absent states and 10 reference-swapped interruption pairs are correct; turn-only resolves 8/10 pairs. These small synthetic diagnostics do not establish recorded-domain or live-interaction performance.

\subsection{Target-Speaker Transcription}
On the original test split, Table~\ref{tab:asr_main} shows that joint prediction reduces target CER from the frozen model's 30.69\% to 4.20\%. Enrollment-free recognizers incur non-target transcription errors; the comparison therefore measures speaker selection and transcription together.

\begin{table}[!t]
\centering\small
\caption{Target-only transcription on the original test split: 350 mixtures (175 per language).}
\label{tab:asr_main}
\setlength{\tabcolsep}{3pt}
\renewcommand{\arraystretch}{1.12}
\begin{tabular*}{\columnwidth}{@{\extracolsep{\fill}}lrrr@{}}\toprule
Model / output & All CER & Zh CER & En WER\\\midrule
FireRedASR2-AED & 72.64 & 68.53 & 78.02\\
Qwen3-ASR-1.7B & 73.63 & 72.97 & 74.91\\
Whisper large-v3 & 73.02 & 72.65 & 75.67\\
CocktailASR-1, native & 30.69 & 44.88 & 29.05\\\midrule
SpeakerTurn, joint & 4.20 & 1.01 & 6.48\\
SpeakerTurn, CoT joint & \textbf{2.56} & \textbf{0.86} & \textbf{4.84}\\
\bottomrule\end{tabular*}
\par\vspace{3pt}{\footnotesize\raggedright CER and WER are percentages. The first three models receive no enrollment. CocktailASR-1 uses its native ASR prompt; SpeakerTurn uses the trained joint prompt. SpeakerTurn omits CoT unless marked.\par}
\end{table}

SpeakerTurn jointly recovers target content and predicts turn states. Its joint transcripts are empty in all 40 target-absent test conditions, consistent with the corresponding absence decisions.

\subsection{Effect of Structured Turn-CoT}
Table~\ref{tab:cot_main} uses the separate validation split under matched update budgets. Its 350 target-present ASR examples (175 per language) are disjoint from the 350 original-test examples in Table~\ref{tab:asr_main}.

\begin{table}[!t]
\centering\small
\caption{Turn detection and target transcription on the separate validation split.}
\label{tab:cot_main}
\setlength{\tabcolsep}{3pt}
\renewcommand{\arraystretch}{1.12}
\begin{tabular*}{\columnwidth}{@{\extracolsep{\fill}}lrr@{}}\toprule
\multicolumn{3}{l}{\textit{(a) Semantic turn detection}}\\
Model / variant & Five F1 & Four F1\\\midrule
CocktailASR-1, frozen$^{\dagger}$ & 0.192 & 0.260\\
SpeakerTurn (no CoT) & \textbf{0.776} & \textbf{0.734}\\
SpeakerTurn (CoT) & 0.762 & 0.716\\\bottomrule\end{tabular*}
\par\vspace{4pt}
\begin{tabular*}{\columnwidth}{@{\extracolsep{\fill}}lrrr@{}}\toprule
\multicolumn{4}{l}{\textit{(b) Target-only transcription}}\\
Model / variant & All CER & Zh CER & En WER\\\midrule
FireRedASR2-AED & 97.26 & 98.89 & 99.60\\
Qwen3-ASR-1.7B & 98.88 & 102.67 & 97.32\\
Whisper large-v3 & 99.05 & 102.79 & 99.16\\
Easy Turn, native & 89.35 & 101.05 & 96.67\\
CocktailASR-1, native & 34.19 & 60.79 & 28.88\\
\midrule
SpeakerTurn (no CoT) & 4.81 & 0.46 & 6.54\\
SpeakerTurn (CoT) & \textbf{3.96} & \textbf{0.44} & \textbf{5.76}\\
\bottomrule\end{tabular*}
\par\vspace{3pt}{\footnotesize\raggedright Five/four-state F1: 420/350 validation conditions; target ASR: the same 350 target-present prefixes for every model. F1 is on $[0,1]$; CER/WER are percentages. All CER pools both languages; Zh CER and En WER use 175 Chinese and 175 English examples, respectively. The first four ASR models receive no enrollment; Easy Turn uses known language codes and omits native control/state tokens. CocktailASR-1 uses native recognition; SpeakerTurn uses joint output. $\dagger$ Group-held-out validation calibration of candidate scores.\par}
\end{table}

CoT improves validation transcription but reduces five- and four-state macro F1, with five-state accuracy unchanged at 84.52\%. The original test confirms this tradeoff: target CER falls from 4.20\% to 2.56\%, whereas five-state macro F1 decreases from 0.811 to 0.764. Backchannel recall drops from 52\% to 32\%, indicating weaker discrimination of brief acknowledgments. The transcription gain therefore does not translate into better turn decisions under this single-seed comparison.

The supplemental set shows the same tradeoff: direct and CoT prediction yield five-state F1 of 0.974 and 0.970, with CER of 1.54\% and 1.18\%, respectively. Joint transcripts are nonempty in 9/750 and 15/750 target-absent cases, contrasting with zero such errors in the 40 original absent conditions. Absence performance therefore varies across these synthetic subsets.

\section{Limitations and Conclusion}
The evaluation is limited by synthetic mixtures, previously observed test data, generator-defined cutoffs, and a single paired training seed. Some backchannel labels appear inconsistent with the corresponding utterance semantics and require independent human review. The reported scores measure agreement with these annotations rather than validated human turn judgments. Controlled noise, reverberation, device echo, continuous endpointing, and generalization to recorded multi-speaker interaction remain unestablished. In addition, structured CoT supervision consists of label-derived templates rather than human reasoning annotations. Adapted models show degraded recognition with the original ASR prompt; the reported SpeakerTurn transcription gains are specific to the trained joint interface.

This study presented SpeakerTurn, an LLM-only adaptation of CocktailASR-1 for joint target transcription and semantic turn detection. The completed no-CoT test achieves five-state macro F1 of 0.811 and joint target CER of 4.20\%. Compared with the earlier turn-only configuration, the joint model provides similar turn performance while supporting accurate transcription through the joint interface. Incomplete-turn and backchannel errors persist. Structured CoT reduces transcription error but lowers turn macro F1 on both validation and test data. Further development should therefore prioritize annotation quality and these semantic distinctions.

\textit{Future work.} We will expand the training corpus and synthesize additional data using a wider range of text-to-speech (TTS) systems to increase speaker, prosodic, and synthesis-style diversity. We will evaluate how data scale and TTS diversity affect target-speaker turn detection and generalization across synthesis systems and recorded speech.

\FloatBarrier

\IEEEtriggeratref{18}

\end{document}